\documentclass[aps,prl,preprint,showpacs,superscriptaddress]{revtex4}

\usepackage{graphicx}
\usepackage{verbatim}
\usepackage{mathrsfs}
\usepackage{gensymb}
\usepackage{color}
\usepackage{ulem}
\usepackage{amsmath,amsfonts,amssymb}
\usepackage{graphicx}

\begin{document}

\title{Multiple Topological Electronic Phases in Superconductor MoC}
\author{Angus~Huang}
\affiliation {Department of Physics, National Tsing Hua University, Hsinchu 30013, Taiwan}
\affiliation {Department of Physics and Astronomy, University of Missouri, Columbia, Missouri 65211, USA}
\author{Adam~D.~Smith}
\affiliation {Department of Physics and Astronomy, University of Missouri, Columbia, Missouri 65211, USA}
\author{Madison~Schwinn}
\affiliation {Department of Physics and Astronomy, University of Missouri, Columbia, Missouri 65211, USA}
\author{Qiangsheng~Lu}
\affiliation {Department of Physics and Astronomy, University of Missouri, Columbia, Missouri 65211, USA}
\author{Tay-Rong~Chang}
\affiliation {Department of Physics, National Cheng Kung University, Tainan, 701, Taiwan}
\author{Weiwei~Xie}
\affiliation {Department of Chemistry, Louisiana State University, Baton Rouge, Louisiana 70803, USA}
\author{Horng-Tay~Jeng\footnote{jeng@phys.nthu.edu.tw}}
\affiliation {Department of Physics, National Tsing Hua University, Hsinchu 30013, Taiwan}
\affiliation {Institute of Physics, Academia Sinica, Taipei 11529, Taiwan}
\author{Guang~Bian\footnote{biang@missouri.edu}}
\affiliation {Department of Physics and Astronomy, University of Missouri, Columbia, Missouri 65211, USA}


\begin{abstract}
The search for a superconductor with non-s-wave pairing is important not only for understanding unconventional mechanisms of superconductivity but also for finding new types of quasiparticles such as Majorana bound states. Materials with both topological band structure and superconductivity are promising candidates as $p+ip$ superconducting states can be generated through pairing the spin-polarized topological surface states. In this work, the electronic and phonon properties of the superconductor molybdenum carbide (MoC) are studied with first-principles methods. Our calculations show that nontrivial band topology and superconductivity coexist in both structural phases of MoC, namely, the cubic $\alpha$ and hexagonal $\gamma$ phases. The $\alpha$ phase is a strong topological insulator and the $\gamma$ phase is a topological nodal line semimetal with drumhead surface states.  In addition, hole doping can stabilize the crystal structure of the $\alpha$ phase and elevate the transition temperature in the $\gamma$ phase. Therefore, MoC in different structural forms can be a practical material platform for studying topological superconductivity and elusive Majorana fermions.


\end{abstract}
\maketitle

\section{Introduction}
The search for topological superconducting materials has sparked tremendous research interest in the solid-state physics community, because topological superconductors host Majorana fermion quasiparticles on their boundaries. The Majorana quasiparticles have been proposed to be the cornerstone for fault-tolerant topological computations \cite{Fisher, Sarma}. One practical route to realizing topological superconductivity is simultaneously combining superconductivity and nontrivial electronic band structure in a single compound. The topological electronic materials, including topological insulators and topological semimetals, possess symmetry-protected surface states \cite{Fu-2011-1, Hsieh-2012-1,Legner-2015-1,Yang-2014-1, Fu-2007-2,Kane-2005-1}, and those surface states usually show nontrivial spin texture in accordance with the nontrivial band topology. For example, the spin and momentum of Dirac surface states in a topological insulator are locked in a helical fashion.  When the topological surface electrons form Cooper pairs in a superconducting state, then an effective time-reversal symmetric $p+ip$ pairing naturally emerges as a consequence of the topological constraint  \cite{Chiu-2016-1,Hasan-2010-1, Fu-2008-1}.
This topological superconducting phase can host Majorana fermions \cite{Bernevig-2013-1,Sato-2017-1} or supersymmetric particles \cite{Xu-2014-1} which have never been experimentally observed in elementary particles. Therefore, it is of fundamental research interest to identify superconductors with nontrivial band topology. 
Several such materials have been predicted and synthesized, $e.g.$, topological insulator/superconductor heterostructures \cite{Xu-2014-1,Fu-2008-1,He-2017-1} and superconducting topological insulators \cite{Chang-2016-1, Guan-2016-1, Bian-2017-1,Wang-2012-1,Hsieh-2012-2,Chen-2016-1}.
Though there has been experimental evidence for Majorana fermions in some of these materials \cite{He-2017-1}, two adverse factors obscure a clear picture of Majorana modes: one is the complex configuration of the heterostructures and the other is low superconducting transition temperature $T_{\mathrm{C}}$, usually below 4K. Therefore, there is a pressing need for the identification of new superconducting compounds with simple composition, high transition temperature and nontrivial electronic band topology.

In this paper, we use first-principles methods to study two structural phases of molybdenum monocarbide (MoC), namely, $\alpha$-MoC and $\gamma$-MoC. The superconductive $\alpha$-MoC phase forms in the rock-salt crystal structure with C deficiency and has a $T_{\mathrm{C}}\simeq14\:\mathrm{K}$ (Ref \cite{Willens-1967-1,Sathish-2014-1}). $\gamma$-MoC, on the other hand, takes a non-centrosymmetric hexagonal structure and is not superconducting in its pristine form. We show that 
$\alpha$-MoC possesses a nonzero $\mathbb{Z}_{2}$ topological invariant and Dirac surface states. $\gamma$-MoC, on the other hand, is a topological nodal line semimetal with drumhead surface states. Moreover, we predict that the $\gamma$ phase of MoC gains superconductivity by hole doping and $T_{\mathrm{C}}$ can be tuned to be higher than 9~K. Therefore, both phases of MoC can be material candidates for studying topological superconductors and their exciting fundamental physics.

\section{Lattice Structure and Bulk Bands}

The $\alpha$ phase of MoC has a rock salt crystal structure with space group 225 ($Fm\bar{3}m$) and the experimental lattice constant is $4.27\:\mathrm{\AA}$ \cite{Clougherty-1961-1}; see Fig.~1(a). The structure is centrosymmetric, and both the Mo atom and the C atom in the unit cell can be regarded as a space inversion center. The first Brillouin zone, a truncated octahedron (same as FCC),  is shown in Fig.~1(b) with the high symmetry points labeled. The bulk band calculation without the inclusion of spin-orbit coupling (SOC) shows that there is a band gap between valence and conduction bands (highlighted in yellow color in Fig.~1(c)) and the gap closes at two points that lie in $X-W$ and $U-X$ as marked in Fig.~1(c). To check these bulk nodal points, we calculated the symmetry properties of the two bands. The space group representation is labeled in the zoomed-in band structure in Figs.~1(e) and 1(f). We find that the two bands that form the bulk band nodes have opposite mirror eigenvalues with respect to the mirror plane $WXU$. Therefore, the intersection of the two bands is protected by the mirror symmetry and the crossing points form a 1D loop surrounding the $X$ point in the $WXU$ plane as schematically drawn in Fig.~1(b). This indicates that $\alpha$-MoC is a nodal line semimetal in the absence of SOC \cite{Burkov2011, Chiu-2016-1, Fang2015, Bian-2016-1, Bian-2016-2}.


Generally, the nodal line band structure is unstable in a centrosymmetric system once SOC is included \cite{Chiu-2016-1}. With space inversion and time reversal symmetries, every band is doubly degenerate with respect to spin, and the two spin subbands carry opposite mirror eigenvalues if the band lies in a mirror plane. So when two spin-degenerate bands cross, the spin subbands with the same mirror eigenvalue repel each other and form a SOC-induced band gap. This is exactly what we find in the band structure of $\alpha$-MoC with SOC taken into consideration; see Figs 1(d, g, h). The bulk nodal line is gapped when SOC is turned on in the calculation. The overall band dispersion changes very little owing to the weak atomic SOC of molybdenum and carbon atoms. The fact that a band crossing is gapped by SOC strongly implies that this phase is a topological insulator. We will show this later in the discussion on the $\mathbb{Z}_{2}$ topological
invariant and the topological surface states of $\alpha$-MoC.

The other structural phase we studied is $\gamma$-MoC which has a noncentrosymmetric hexagonal structure (space group 187, $P\bar{6}m2$) like tungsten monocarbide (WC); see Fig.~1(i). The lattice constants are $a=2.898\:\mathrm{\AA}$ and $c=2.809\:\mathrm{\AA}$ \cite{Kuo-1952-1}.  The WC structure is comprised of hexagonal atomic layers which stack along the (001) direction. The Mo layer is sandwiched between two C layers in an ABA fashion, and this trilayer structure forms the unit cell of $\gamma$-MoC. The first Brillouin zone and the high symmetry points of WC structure are depicted in Fig. 1(j). The calculated bulk bands with and without SOC are shown in Figs.~1(k) and 1(l). The bulk band gap is highlighted in yellow color. In the spinless case, there are two gapless features near the Fermi level.  One is a bulk nodal ring (loop-shaped nodal line) surrounding the $K$ point. Similar to the case of spinless $\alpha$-MoC, the nodal ring is protected by the mirror symmetry with respect to the $\Gamma MK$ plane, which can be seen from mirror parity eigenvalues labeled in Fig.~1(m). The other feature is a band crossing of a two-fold degenerate band (the red solid and dashed lines in Fig.~1(k)) and a single spinless band along the $\Gamma-A$ direction. The two-fold degenerate band is a two-component representation of the "little" space group in the $\Gamma-A$ direction. Thus, the degeneracy of the crossing point marked by a red circle in Fig.~1(k) is three, which was referred to as  ``class $\alpha$ 3-fold fermion" in the literature \cite{Chang-2017-1}. The band plotted with a red solid line can be regarded as a bulk nodal line in the form of an open curve.  When SOC is included, each spinless nondegenerate band splits into two spinful subbands due to the lack of space inversion symmetry while the spinless double band transforms into two spinful double bands as shown in Figs. 1(l, n, o). The spinless nodal ring now becomes two copies of Weyl nodal rings which remain gapless due to the protection of mirror reflection symmetry \cite{Bian-2016-1, Bian-2016-2, Sun-2017-1}; see Fig.~1(n).  Along $\Gamma-A$, four 3-fold fermion nodes form due to the band splitting. They are connected by two doubly degenerate nodal lines as demonstrated in Fig. 1(o) \cite{Chang-2017-1,Sun-2017-1}. Here we note the difference between the
$\Gamma-A$ nodal line and the nodal ring surrounding the $K$
point. The nodal lines connecting two 3-fold fermion nodes are protected
by the 3-fold rotational symmetry along $\Gamma-A$ while the nodal ring is protected
by the mirror reflection symmetry. 

\section{Topological Invariant and Surface Band Structure}

The topological invariant and surface bands of the two phases of MoC are calculated and the results show that both phases are topologically nontrivial in their bulk and surface band structure; see Figs.~2 and 3. The nodal ring of $\alpha$-MoC is gapped by SOC, resulting in a continuous band gap that traverses the whole Brillouin zone. The crystal lattice of $\alpha$-MoC has space inversion symmetry, so the $\mathbb{Z}_{2}$ topological invariant can be determined by the parity eigenvalues of all valence bands at the eight time-reversal invariant momentum (TRIM) points which are $\Gamma$(1), $X$(3), $L$(4) for $\alpha$-MoC. According to our calculations, the product of parity eigenvalues  is $-1$, 1, and 1 for $\Gamma$, $X$, and $L$, respectively. Therefore, the $\mathbb{Z}_{2}$ topological invariant is $\nu_{\mathbb{Z}2}=\prod_{i\in \mathrm{TRIM}}P_{i}=-1$, indicating $\alpha$-MoC is a strong topological insulator \cite{Fu-2007-1}. We also calculated the $\mathbb{Z}_{2}$ topological invariant by the Wilson loop method \cite{Yu2011}; see Fig. 2(f). The Wilson band is an open curve traversing the entire Brillouin zone in the time reversal invariant plane $k_{z}=0$ and a closed loop in another time reversal invariant plane $k_{z}=0.5\pi$. The result indicates that the $\mathbb{Z}_{2}$ invariant equals 1, which is consistent with the result from our parity calculations.

The surface states of  $\alpha$-MoC  are calculated in a semi-infinite slab geometry with (111) surface termination. The bulk and surface bands of $\alpha$-MoC (111) with and without SOC are plotted in Figs.~2(a) and 2(b). In the case without SOC, the bulk nodal ring (NR) surrounds the $\bar{M}$ point which corresponds to the projection of $X$ to the (111) surface Brillouin zone. The 2D drumhead surface band  connects to the bulk band nodes inside the ring as marked in Fig.~2(a). There is an extra surface band lying outside the nodal ring along $\bar{M}$-$\bar{K}$. When SOC is included, the bulk nodal ring is gapped, and the drumhead surface band transforms into topological Dirac surface states in a strong topological insulator. We note that the surface band outside the nodal ring splits into two subbands, one of which is connected to the lower branch of the Dirac cone as shown in the zoomed-in bands in Fig.~2(c). With this connection to the outer surface band, the Dirac cone extends to the valence bulk band along $\bar{M}-\bar{K}-\bar{\Gamma}$ and form a gapless dispersion connecting the conduction and valence bulk bands as required by the band topology. The dispersion of the surface bands in the presence of SOC again proves that $\alpha$-MoC is a strong topological insulator. The 2D iso-energy contours are plotted in Figs.~2(d) and 2(e) for the cases without and with SOC, respectively. The energy is set to 1.7 eV, which is the energy value of the bulk node in the $\Gamma-M$ direction. The nodal ring, drumhead surface states and topological Dirac surface states can be seen in the contour. Interestingly, the surface state outside the nodal ring form a surface arc with both its ends connected to the bulk electron pocket as shown in Fig.~2(d).

There is no absolute band gap in $\gamma$-MoC, so the $\mathbb{Z}_{2}$ topological invariant can not be defined in this phase. Instead, we can calculate the winding number associated with the nodal ring around the K point. The winding number has been shown to be equal to a mirror Chern number $n^{+}$ which is defined as the difference between the number of valence bands with a +1 mirror
eigenvalue inside and outside the nodal ring \cite{Bian-2016-1}. $n^{+}= -1$ for the spinless nodal ring according to the result in Fig 1(n). In the presence of SOC, $n^{+}=\pm1$ for inner and outer
nodal rings, respectively. The winding number is a topological invariant and it determines the number of drumhead surface bands emanating from the nodal ring. To show this, we calculated the surface bands of $\gamma$-MoC in a semi-infinite slab geometry with the (001) surface. There are two possible terminations: one is with a C top layer and the other with a Mo top layer.  In both cases, we found a single nodal ring surrounding the $\bar{K}$ in the bulk band and a drumhead surface band which disperses outwards from the ring in the absence of SOC; see Fig.~3. When SOC is considered, the nodal ring splits into two copies of Weyl nodal rings and each Weyl ring is connected to a spinful drumhead surface band \cite{Bian-2016-1}. The drumhead bands in the cases of C and Mo terminations have opposite signs in their slope outside the ring, which can be attributed to the dramatically different atomic electronegativity of C and Mo surface layers.
%
%

\section{Electron-Phonon Coupling and Superconductivity}

To investigate the superconductivity and electron-phonon coupling (EPC) properties
of the two phases of MoC, We performed a systematic density functional perturbation therory (DFPT) simulation (see method section for details). We found that $\alpha$-MoC is unstable in its pristine form and the phonon bands have imaginary frequencies, which is consistent with previous theoretical studies \cite{Isaev-2007-1,Isaev-2005-1, Hart-2000-1}. In experiments, the $\alpha$ phase of MoC has been synthesized but with C deficiency (MoC$_{1-\delta}$)\cite{Athanasiou-1997-1,Sathish-2014-1}. The C deficiency effectively lowers the Fermi level since the total number of electrons per formula unit decreases. To simulate the effect of C deficiency, we calculated the EPC and phonon properties with hole doping; see Fig.~4(a). We found $\alpha$-MoC is indeed unstable in its pristine form. The phonon bands have imaginary frequencies, consistent with previous theoretical studies. When the hole doping increases to be  more than $-0.2$ electron per formula unit ($-0.2\:\mathrm{e/fu}$), the imaginary phonon band disappears (Fig 4.c) and the lattice is stabilized. This suggests that $\alpha$-MoC with hole doping can be a stable superconducting phase, which is in accordance with the experimental results. We note that previous studies suggest that the Coulomb repulsion $\mu^{*}$ may increase in transition metal carbides (TMC) and transition metal nitride (TMN) \cite{Isaev-2007-1,Isaev-2005-1}. This occurs in disordered structures because defects can enhance the electron-electron
Coulomb interaction \cite{Szabo-2016-1}. In our simulations, we set $\mu^{*}=0.15$, which is larger than the commonly adopted value of 0.1, to take into account the defect effect. The transition temperature is $T_{\mathrm{C}}\simeq21.1\:\mathrm{K}$ at doping level $-0.2\:\mathrm{e/fu}$,
and decreases gradually with the increasing concentration of of hole doping. $T_{\mathrm{C}}=15.7\:\mathrm{K}$ for $-0.5\:\mathrm{e/fu}$ hole doping, which is similar
to the experimental value of 14.3$\:\mathrm{K}$ \cite{Willens-1967-1}. Further increasing hole doping to $-1.0\:\mathrm{e/fu}$ brings $T_{C}$ down to 10.4$\:\mathrm{K}$. This is consistent with the experimental results from $\mathrm{NbC}$ ($T_{\mathrm{C}}\simeq11.1\:\mathrm{K}$) \cite{Willens-1967-1}, since Nb has one less electron compared to Mo. The trend
of $T_{C}$ with varied doping agrees well with the experiment values obtained from $\mathrm{MoC}-\mathrm{NbC}$ alloy \cite{Willens-1967-1}.

In the case without doping, one of acoustic phonon bands of $\alpha$-MoC becomes imaginary near the $X$ point; see Fig.~4(b). The EPC strength (represented d by the red line thickness) diverges as the phonon energy approaches zero along $\Gamma$-$X$. Both facts indicate that the crystal structure of $\alpha$-MoC is unstable. Doping holes into the compound can remove the imaginary phonon modes and create soft acoustic modes around $X$. Those soft phonon modes couple strongly with electronic states and thus dominantly contribute to the effective EPC in the BCS theory; see Fig.~4(c). Therefore, the structural instability of $\alpha$-MoC is responsible for the high transition temperature observed in this phase. The close relationship of superconductivity and structural instability has also been found in materials with charge density waves (CDW). Suppressing CDW may generate superconductivity in CDW materials, such as in $\mathrm{Cu}_{x}\mathrm{Ta}\mathrm{S}_{2}$ \cite{Wagner-2008-1} and $\mathrm{PbTa}\mathrm{Se}_{2}$ \cite{Chang-2016-1}.  As the hole doping concentration increases, the soft phonon modes become normal phonon modes. At the same time, the effective EPC becomes weaker and consequently the superconducting transition temperature $T_{\mathrm{C}}$ decreases as shown in Fig. 4(a).

%
%

Unlike $\alpha$-MoC, the crystal structure of  $\gamma$-MoC is stable in its pristine form but superconductivity has not been observed in $\gamma$-MoC. We calculated the transition temperature of $\gamma$-MoC with various doping concentrations and the result is plotted in Fig.~4(d).$T_{\mathrm{C}}$ is negligibly small with or without electron doping. Hole doping, on the other hand, can induce superconductivity in $\gamma$-MoC.
Doping 0.3 hole per formula unit can lift $T_{\mathrm{C}}$ to 0.4$\:\mathrm{K}$, and $T_{\mathrm{C}}$ increases rapidly as the hole doping concentration further increases. For example,$T_{\mathrm{C}}\simeq8.6\:\mathrm{K}$ at hole doping level $-0.7\:\mathrm{e/fu}$.

The phonon band and EPC of $\gamma$-MoC with different levels of hole doping are shown
in Figs. 4(e) and 4(f). The  phonon dispersion only changes slightly by increasing the doping level, but the EPC is remarkably enhanced for almost all phonon bands. The enhancement of EPC and superconductivity in $\gamma$-MoC by hole doping can be attributed to the increase in the electron density of states (DOS), especially when the Fermi level passes the Van Hove singularity of DOS at about 0.8 eV below the Fermi level at $H$ and $A$ points. Thus we predict that hole doped $\gamma$-MoC is a superconducting nodal-line semimetal with drumhead surface states.


\section{Conclusion}

In summary, we performed a systematic first-principles study of the two structural phases of MoC. $\alpha$-MoC is a superconductor with a considerably high T$_c$ of 14~K. The pristine rocksalt lattice of $\alpha$-MoC is unstable according to our phonon calculations. However, it can be stabilized by hole doping, which is in line with the fact that the chemically synthesized $\alpha$-MoC carries carbon vacancies. The high transition temperature of $\alpha$-MoC is connected with the suppression of structural instability. Moreover, $\alpha$-MoC is a strong topological insulator with a nonzero topological $\mathbb{Z}_{2}$ invariant. The Dirac surface band is located at 1.7~eV above the Fermi level. $\gamma$-MoC, on the other hand, is not superconducting in its pristine form. Hole doping can make the compound a superconductor. The transition temperature can be as high as 9 K if 0.75 hole is doped per unit cell.  We also found that $\gamma$-MoC is a topological nodal-line semimetal with a pair of nodal rings surrounding the K point and drumhead surface states. Besides MoC, transition metal monocarbides, XC (X = W, Ta, Nb), and transition metal mononitrides, XN (X = V, Nb, Ti), are all superconductors \cite{Webb_SC, Meissner-1930-1}, and meanwhile they can host a large variety of band topology in their different structural forms. Therefore, these compounds provide an exciting family of candidate materials for studying the rich physics in topological superconductors and Majorana bound states.

\section{Methods}

The first-principles electronic structure simulations are based on the
density functional theory (DFT) which is implemented in VASP packages \cite{Kresse-1993-1,Kresse-1996-1,Kresse-1996-2}.
The PBE-type \cite{Perdew-1996-1} GGA functionals and PAW-type pseudopotential \cite{Kresse-1999-1,Blochl-1994-1} are used in the calculations. The
lattice parameters are taken from experimental values.
A $24\times24\times24$ $k$ grid and an energy cutoff of $400\:\mathrm{eV}$
are used in the band structure calculations. The vasp2wannier90 interface \cite{Franchini-2012-1} is used to construct the Wannier Hamiltonion of semi-infinite slabs. The surface band spectrum is obtained from the Green function of semi-infinite slabs \cite{Zhang-2009-1}. The topological
$\mathbb{Z}_{2}$ invariant is calculated from
the phase of Wannier centers \cite{Soluyanov-2011-1}. 

The calculations of electron-phonon coupling and superconductivity are
based on the density functional perturbation theory (DFPT). Quantum
Espresso packages are employed with norm-conserving pseudopotentials in the phonon calculations. 
$24\times24\times24$ ($30\times30\times32$) $k$ grid and $4\times4\times4$ ($3\times3\times4$) $q$ grid
are used for $\alpha$ phase ($\gamma$ phase) MoC in the DFPT calculations. The energy cutoff is 40 Ry (400 Ry) for wave function (charge density) calculations. $0.02\:\mathrm{Ry}$ broadening
is set in the Fermi-Dirac distribution for phonon mode calculations.  The total electron-phonon coupling is given by

\begin{equation}
\lambda=\sum_{\mathbf{q}v}\lambda_{\mathbf{q}v}
\end{equation}
where $\lambda_{\mathbf{q}v}$ is the electron phonon
coupling strength for different phonon state,

\begin{equation}
\lambda_{\mathbf{q}v}\simeq\frac{1}{N_{\mathrm{F}}\omega_{\mathbf{q}v}}\sum_{mn,\mathbf{k}}\Bigl|g_{mn}^{v}(\mathbf{k},\mathbf{q})\Bigr|^{2}\delta(\varepsilon_{n\mathbf{k}})\delta(\varepsilon_{m\mathbf{k+q}}).
\end{equation}
$g_{mn}^{v}$ is the dynamical matrix and $N_{\mathrm{F}}$
is the density of states at the Fermi level for a single spin. $\omega_{\mathbf{q}v}$ and $\varepsilon_{n\mathbf{k}}$ are the energy eigenvalues of phonon bands
and electron bands,
respectively. The transition temperature $T_{\mathrm{C}}$ is calculated by using the McMillan
formula \cite{McMillan-1968-1,Allen-1975-1},

\begin{equation}
T_{\mathrm{C}}\simeq\frac{\bar{\omega}_{\mathrm{log}}}{1.2}\exp\left(\frac{1.04\left(1+\lambda\right)}{\mu^{*}\left(1+0.62\lambda\right)-\lambda}\right).
\end{equation}
where $\mu^{*}$ is the Morel-Anderson pseudopotential \cite{Morel-1962-1},
and is usually set between 0.1 and 0.2. It describes the effective electron-electron Coulomb repulsion. $\bar{\omega}_{\mathrm{log}}$ describes the average
of phonon frequency. To simulate the charge doping, we change the
total number of electrons in one unit cell artificially and insert a uniform charge background to keep the system charge neutral.
The self-consistent DFT and DFPT calculations are repeated for each
doping concentration.

\newpage{}

\begin{figure}
\begin{centering}
\includegraphics[width=16cm]{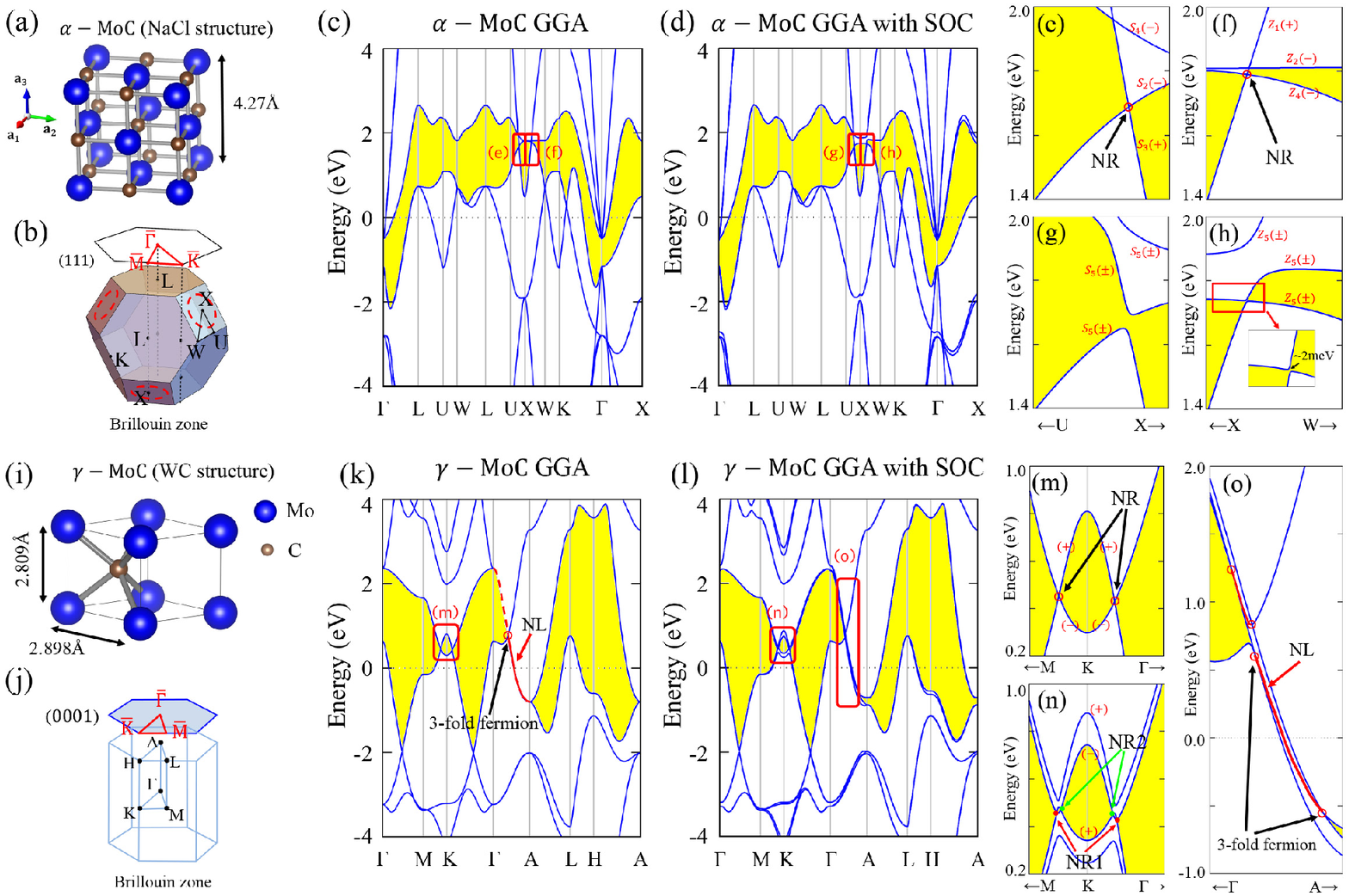}
\par\end{centering}
\caption{ (a) The lattice structure of $\alpha$-MoC. The Mo and C atoms are represented by blue and brown balls, respectively. (b) The first Brillouin zone and (111)-surface Brillouin zone of $\alpha$-MoC (c, d) The bulk band structure of $\alpha$-MoC without and with SOC, respectively. A continuous bulk gap is highlighted in yellow color. (e-h) The zoomed-in band crossings marked in (a, b). The symmetry (space group representation) are labeled and the mirror eigenvalue in the $XUW$ mirror plane is given by $+$ or $-$ sign. In (e, f) The nodal ring (NR) surrounding $X$ are highlighted by red circles. The band crossing is protected by the mirror symmetry.  (i-l) Same as (a-d), but for $\gamma$-MoC. In (j), the (001) surface Brillouin zone is plotted. (m, n) The zoomed-in nodal ring band structure surrounding the $K$ point marked in (k, l), respectively. The mirror eigenvalue in the $\Gamma KM$ plane is labeled by $+$ or $-$ sign. (o) The zoom-in band structure of $\gamma$-MoC between $\Gamma A$ as marked in (l). The nodal lines (NL, red lines) connect 3-fold degenerate bulk nodes (red circles).}
\end{figure}

\newpage{}

\begin{figure}
\begin{centering}
\includegraphics[width=16cm]{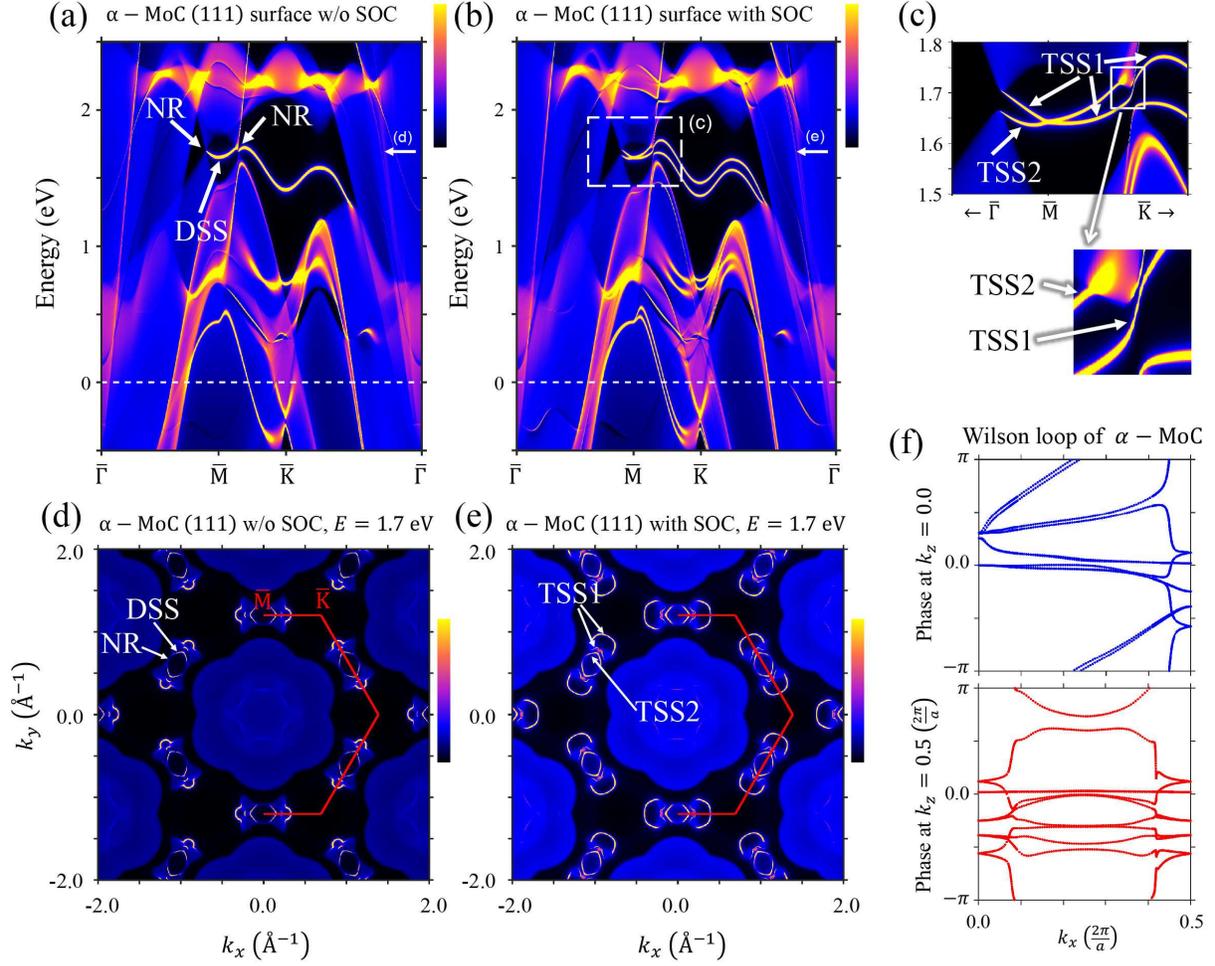}
\par\end{centering}
\caption{(a, b)  Band spectrum of semi-infinite $\alpha$-MoC without and with the inclusion of SOC, respectively. The bands are colored according to the surface weight of each state. (c) The zoomed-in topological surface band structure marked in (b). (d, e) The iso-energy band contour at $E=1.7$ eV without and with SOC, respectively. The energy value is marked by white arrows in (a, b). (f) The calculated Wilson loop showing a non-zero topological $\mathbb{Z}_{2}$ invariant. }

\end{figure}

\newpage{}

\begin{figure}
\begin{centering}
\includegraphics[width=16cm]{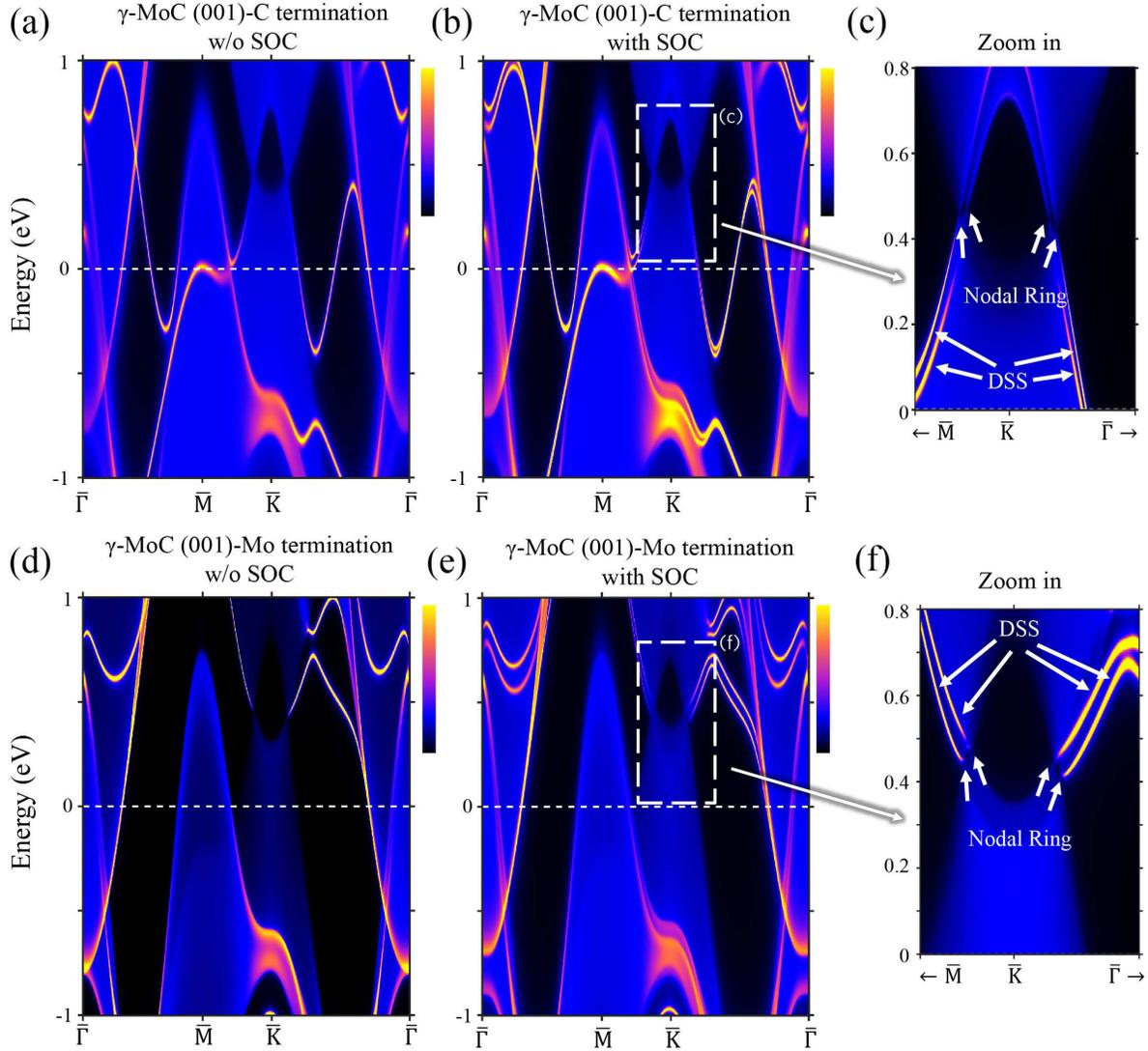}
\par\end{centering}
\caption{(a, b) Band spectrum of semi-infinite $\gamma$-MoC without and with the inclusion of SOC, respectively. The bands are colored according to the surface weight of each state. The surface is terminated in (001) direction with a C top layer. (c) The zoomed-in nodal line band structure marked in (b). (d-f) Same as (a-c), but for a Mo-terminated (001) surface.}

\end{figure}

\newpage{}

\begin{figure}
\begin{centering}
\includegraphics[width=16cm]{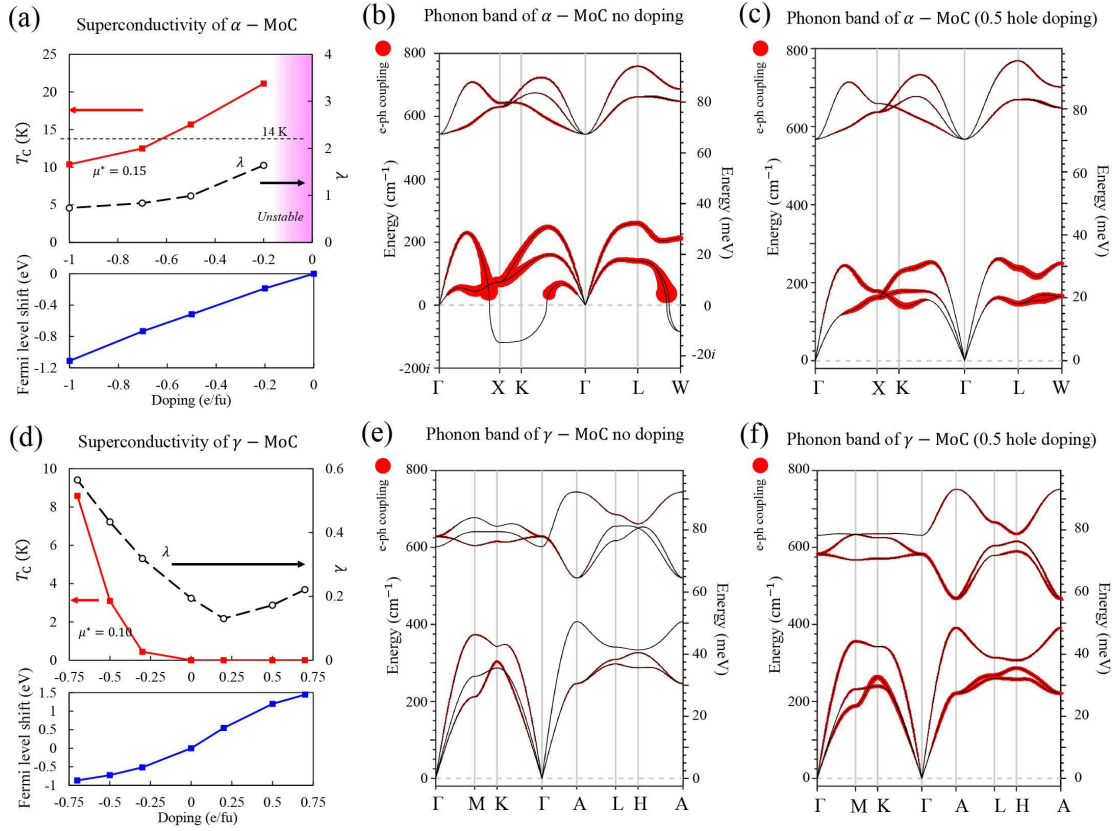}
\par\end{centering}
\caption{Calculated phonon dispersion and superconducting transition temperature of MoC in two phases. (a) The transition
temperature (red line) and electron-phonon coupling
strength $\lambda$ (black dash line) of $\alpha$-MoC as a function of doping concentration (negative values correspond to hole doping).
The energy shift of Fermi level is plotted the in lower
subfigure. (b, c) The phonon bands of the $\alpha$-MoC without doping and with $-0.5\:\mathrm{e/fu}$ doping, respectively. The thickness of red line indicates the strength of electron-phonon coupling.
(d-f) Same as (a-c), but for $\gamma$-MoC. }
\end{figure}


%
%
%
%

\end{document}